\RequirePackage{lineno}
\documentclass[aps,prl,showpacs,twocolumn,superscriptaddress,groupedaddress,floatfix,amsfonts]{revtex4}
\usepackage{graphicx}
\usepackage{dcolumn}
\usepackage{bm}
\usepackage{amssymb}
\usepackage{amsmath}
\usepackage{xspace}
\newcommand{\la}{\langle}
\newcommand{\ra}{\rangle}
\newcommand{\lt}{\!<\!}

\newcommand{\GeV}{\ensuremath{\text{GeV}}\xspace}
\newcommand{\TeV}{\ensuremath{\text{TeV}}\xspace}
\newcommand{\Ptg}{p_{T}^{\gamma}\xspace}
\newcommand{\ptg}{\ensuremath{p_T^{\gamma}}\xspace}
\newcommand{\gb}{\ensuremath{\gamma+{b}}\xspace}

\usepackage{color}
\newcommand{\CHECK}[1]{\textbf{\color{red}[#1]}\xspace}
\newcommand{\chk}[1]{\CHECK{CHECK THIS!}}

\begin{document}

\hspace{5.2in} \mbox{FERMILAB-PUB-12-082-E}

\title{\boldmath Measurement of the photon$+b$-jet production 
differential cross section in $p\bar{p}$ collisions at $\sqrt{s}=1.96~\TeV$}

\affiliation{LAFEX, Centro Brasileiro de Pesquisas F\'{i}sicas, Rio de Janeiro, Brazil}
\affiliation{Universidade do Estado do Rio de Janeiro, Rio de Janeiro, Brazil}
\affiliation{Universidade Federal do ABC, Santo Andr\'e, Brazil}
\affiliation{University of Science and Technology of China, Hefei, People's Republic of China}
\affiliation{Universidad de los Andes, Bogot\'a, Colombia}
\affiliation{Charles University, Faculty of Mathematics and Physics, Center for Particle Physics, Prague, Czech Republic}
\affiliation{Czech Technical University in Prague, Prague, Czech Republic}
\affiliation{Center for Particle Physics, Institute of Physics, Academy of Sciences of the Czech Republic, Prague, Czech Republic}
\affiliation{Universidad San Francisco de Quito, Quito, Ecuador}
\affiliation{LPC, Universit\'e Blaise Pascal, CNRS/IN2P3, Clermont, France}
\affiliation{LPSC, Universit\'e Joseph Fourier Grenoble 1, CNRS/IN2P3, Institut National Polytechnique de Grenoble, Grenoble, France}
\affiliation{CPPM, Aix-Marseille Universit\'e, CNRS/IN2P3, Marseille, France}
\affiliation{LAL, Universit\'e Paris-Sud, CNRS/IN2P3, Orsay, France}
\affiliation{LPNHE, Universit\'es Paris VI and VII, CNRS/IN2P3, Paris, France}
\affiliation{CEA, Irfu, SPP, Saclay, France}
\affiliation{IPHC, Universit\'e de Strasbourg, CNRS/IN2P3, Strasbourg, France}
\affiliation{IPNL, Universit\'e Lyon 1, CNRS/IN2P3, Villeurbanne, France and Universit\'e de Lyon, Lyon, France}
\affiliation{III. Physikalisches Institut A, RWTH Aachen University, Aachen, Germany}
\affiliation{Physikalisches Institut, Universit\"at Freiburg, Freiburg, Germany}
\affiliation{II. Physikalisches Institut, Georg-August-Universit\"at G\"ottingen, G\"ottingen, Germany}
\affiliation{Institut f\"ur Physik, Universit\"at Mainz, Mainz, Germany}
\affiliation{Ludwig-Maximilians-Universit\"at M\"unchen, M\"unchen, Germany}
\affiliation{Fachbereich Physik, Bergische Universit\"at Wuppertal, Wuppertal, Germany}
\affiliation{Panjab University, Chandigarh, India}
\affiliation{Delhi University, Delhi, India}
\affiliation{Tata Institute of Fundamental Research, Mumbai, India}
\affiliation{University College Dublin, Dublin, Ireland}
\affiliation{Korea Detector Laboratory, Korea University, Seoul, Korea}
\affiliation{CINVESTAV, Mexico City, Mexico}
\affiliation{Nikhef, Science Park, Amsterdam, the Netherlands}
\affiliation{Radboud University Nijmegen, Nijmegen, the Netherlands}
\affiliation{Joint Institute for Nuclear Research, Dubna, Russia}
\affiliation{Institute for Theoretical and Experimental Physics, Moscow, Russia}
\affiliation{Moscow State University, Moscow, Russia}
\affiliation{Institute for High Energy Physics, Protvino, Russia}
\affiliation{Petersburg Nuclear Physics Institute, St. Petersburg, Russia}
\affiliation{Instituci\'{o} Catalana de Recerca i Estudis Avan\c{c}ats (ICREA) and Institut de F\'{i}sica d'Altes Energies (IFAE), Barcelona, Spain}
\affiliation{Uppsala University, Uppsala, Sweden}
\affiliation{Taras Shevchenko National University of Kyiv, Kiev, Ukraine}
\affiliation{Lancaster University, Lancaster LA1 4YB, United Kingdom}
\affiliation{Imperial College London, London SW7 2AZ, United Kingdom}
\affiliation{The University of Manchester, Manchester M13 9PL, United Kingdom}
\affiliation{University of Arizona, Tucson, Arizona 85721, USA}
\affiliation{University of California Riverside, Riverside, California 92521, USA}
\affiliation{Florida State University, Tallahassee, Florida 32306, USA}
\affiliation{Fermi National Accelerator Laboratory, Batavia, Illinois 60510, USA}
\affiliation{University of Illinois at Chicago, Chicago, Illinois 60607, USA}
\affiliation{Northern Illinois University, DeKalb, Illinois 60115, USA}
\affiliation{Northwestern University, Evanston, Illinois 60208, USA}
\affiliation{Indiana University, Bloomington, Indiana 47405, USA}
\affiliation{Purdue University Calumet, Hammond, Indiana 46323, USA}
\affiliation{University of Notre Dame, Notre Dame, Indiana 46556, USA}
\affiliation{Iowa State University, Ames, Iowa 50011, USA}
\affiliation{University of Kansas, Lawrence, Kansas 66045, USA}
\affiliation{Kansas State University, Manhattan, Kansas 66506, USA}
\affiliation{Louisiana Tech University, Ruston, Louisiana 71272, USA}
\affiliation{Boston University, Boston, Massachusetts 02215, USA}
\affiliation{Northeastern University, Boston, Massachusetts 02115, USA}
\affiliation{University of Michigan, Ann Arbor, Michigan 48109, USA}
\affiliation{Michigan State University, East Lansing, Michigan 48824, USA}
\affiliation{University of Mississippi, University, Mississippi 38677, USA}
\affiliation{University of Nebraska, Lincoln, Nebraska 68588, USA}
\affiliation{Rutgers University, Piscataway, New Jersey 08855, USA}
\affiliation{Princeton University, Princeton, New Jersey 08544, USA}
\affiliation{State University of New York, Buffalo, New York 14260, USA}
\affiliation{Columbia University, New York, New York 10027, USA}
\affiliation{University of Rochester, Rochester, New York 14627, USA}
\affiliation{State University of New York, Stony Brook, New York 11794, USA}
\affiliation{Brookhaven National Laboratory, Upton, New York 11973, USA}
\affiliation{Langston University, Langston, Oklahoma 73050, USA}
\affiliation{University of Oklahoma, Norman, Oklahoma 73019, USA}
\affiliation{Oklahoma State University, Stillwater, Oklahoma 74078, USA}
\affiliation{Brown University, Providence, Rhode Island 02912, USA}
\affiliation{University of Texas, Arlington, Texas 76019, USA}
\affiliation{Southern Methodist University, Dallas, Texas 75275, USA}
\affiliation{Rice University, Houston, Texas 77005, USA}
\affiliation{University of Virginia, Charlottesville, Virginia 22901, USA}
\affiliation{University of Washington, Seattle, Washington 98195, USA}
\author{V.M.~Abazov} \affiliation{Joint Institute for Nuclear Research, Dubna, Russia}
\author{B.~Abbott} \affiliation{University of Oklahoma, Norman, Oklahoma 73019, USA}
\author{B.S.~Acharya} \affiliation{Tata Institute of Fundamental Research, Mumbai, India}
\author{M.~Adams} \affiliation{University of Illinois at Chicago, Chicago, Illinois 60607, USA}
\author{T.~Adams} \affiliation{Florida State University, Tallahassee, Florida 32306, USA}
\author{G.D.~Alexeev} \affiliation{Joint Institute for Nuclear Research, Dubna, Russia}
\author{G.~Alkhazov} \affiliation{Petersburg Nuclear Physics Institute, St. Petersburg, Russia}
\author{A.~Alton$^{a}$} \affiliation{University of Michigan, Ann Arbor, Michigan 48109, USA}
\author{G.~Alverson} \affiliation{Northeastern University, Boston, Massachusetts 02115, USA}
\author{M.~Aoki} \affiliation{Fermi National Accelerator Laboratory, Batavia, Illinois 60510, USA}
% ...
%
%
\collaboration{The D0 Collaboration\footnote{with visitors from
$^{a}$Augustana College, Sioux Falls, SD, USA,
$^{b}$The University of Liverpool, Liverpool, UK,
$^{c}$UPIITA-IPN, Mexico City, Mexico,
$^{d}$DESY, Hamburg, Germany,
,
$^{e}$SLAC, Menlo Park, CA, USA,
$^{f}$University College London, London, UK,
$^{g}$Centro de Investigacion en Computacion - IPN, Mexico City, Mexico,
$^{h}$ECFM, Universidad Autonoma de Sinaloa, Culiac\'an, Mexico
and
$^{i}$Universidade Estadual Paulista, S\~ao Paulo, Brazil.
$^{\ddag}$Deceased.
}} \noaffiliation
\vskip 0.25cm

\date{\today}

\begin{abstract}
  We present measurements of the differential cross section ${\rm d}\sigma/{\rm d}\Ptg$
  for the inclusive production of a photon in association with a $b$-quark jet
  for photons with rapidities $|y^\gamma|\lt 1.0$ and  
  $30<\ptg <300$~\GeV, as well as for photons with $1.5<|y^\gamma|\lt 2.5$
  and $30<\ptg <200$~\GeV, where $\Ptg $ is the photon transverse momentum. 
  The $b$-quark jets are required to have $p_T>15$ GeV and rapidity $| y^\text{jet}|\lt 1.5$.
  The results are based on data corresponding to an integrated luminosity of 8.7 fb$^{-1}$,
  recorded with the D0 detector at the Fermilab Tevatron $p\bar{p}$ Collider at $\sqrt{s}=$1.96~\TeV.
  The measured cross sections are compared with next-to-leading order perturbative QCD calculations 
  using different sets of parton distribution functions as well as to predictions
  based on the $k_{\rm T}$-factorization QCD approach, and those from the {\sc sherpa} and {\sc pythia} Monte Carlo event generators.

\end{abstract}

\maketitle



In hadron-hadron collisions, high-energy photons ($\gamma$)
emerge unaltered from the hard scattering process of two partons and 
therefore provide a clean probe of the parton level dynamics. 
Study of such photons (called prompt or direct) produced in association with a $b$-quark 
also provides information about the $b$-quark and gluon ($g$) parton distribution functions (PDFs) of the
incoming hadrons.
Such events are produced in Quantum Chromodynamics (QCD) primarily through the Compton-like scattering process
$gb\to \gamma b$, which dominates up to photon transverse momenta
($\Ptg$) of $\approx 70$~GeV,
and through quark-antiquark annihilation 
$q\bar{q}\to \gamma g \to \gamma b\bar{b}$, which dominates at high $\Ptg$~\cite{Tzvet}.
The inclusive $\gb$ production may also 
originate from partonic processes like $gg \to b\bar{b}$ or $bg \to bg$,
where the final state $b$-quark or gluon fragments into a photon \cite{Tzvet}. 
However, photon isolation requirements 
substantially reduce the 
contributions from this process.
The measurements of the differential cross section as a function
of $\ptg$ 
and the photon (and/or $b$-jet) rapidity
can be used  to test the \gb production mechanism and  
the underlying dynamics of QCD hard-scattering subprocesses 
with different momentum transfer scales $Q$ and parton momentum fraction $x$.
Measurements involving $\gamma/Z$-boson and $b$-quark final states
have previously been performed by the D0 and CDF collaborations  
\cite{Zb,Z_b_CDF,Z_b_D0,gamma_b_cdf,gamma_b_d0}.
In comparison to the previous \gb measurement~\cite{gamma_b_d0},
we now consider not only the leading (in $p_T$) $b$-jet, but all $b$-jets in the event.
To increase statistics in $\Ptg$ bins, we have also extended 
the $|y^{\rm jet}|$ region which results in a larger contribution from
the annihilation process.
The large integrated luminosity recorded with the D0
detector in $p\bar{p}$ collisions at $\sqrt{s} = 1.96$ TeV at the Fermilab Tevatron Collider 
and more advanced photon and $b$-jet identification tools~\cite{diphoton,hgg_prl,b-NIM}
enable us to perform more precise measurements and to extend them in kinematic regions
previously unexplored.

\begin{figure}[thbp]
  \centering
  \includegraphics[width=0.8\linewidth]{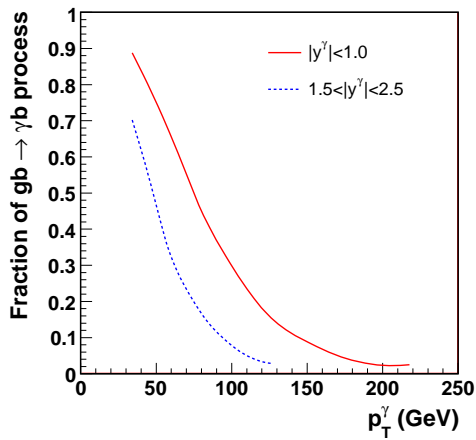}
  \vspace{-4mm}
  \caption{Fractional contribution of the $gb\to \gamma b$ subprocess to
   the associated production of direct photon and $b$-jet
   as a function of $\Ptg$
   in the events with photons in the central and forward rapidity regions.
   The fractions are calculated using {\sc pythia} 6.4~\cite{PYT} and the {\sc cteq}6.1L parton distribution functions~\cite{CTEQ}.
    }
  \label{fig:qg_frac}
\end{figure}

In this Letter, we present measurements of the inclusive $\gamma+b$-jet
production cross sections using data collected from June 2006 to September 2011. 
The cross sections are measured as a function of
$\Ptg$ in the photon rapidity regions, $|y^\gamma|\lt 1.0$ (central)
and $1.5<|y^\gamma|\lt 2.5$ (forward). 
The rapidity, $y$, is related to the polar scattering 
angle $\theta$ with respect to the proton beam axis by
$y = \frac{1}{2} \ln[(1 + \beta cos\theta)/(1 −- \beta cos\theta)]$, 
where $\beta$ is defined as the ratio between momentum 
and energy $\beta = |\vec{p}|/E$.
The photons are required to have $30<\Ptg< 300$~GeV in the central rapidity region
and  $30<\Ptg< 200$~GeV in the forward region.
The $b$-jets are required to be within $| y^\text{jet}|\lt 1.5$ and to
have transverse momentum $p_{T}^{\rm jet}>15$~GeV.
This allows us to probe the dynamics of the production process in 
a wide kinematic range, not studied before in other measurements of 
a vector boson + $b$-jet final state. The measurement covers
parton momentum fractions in the range $0.007 \lesssim x \lesssim 0.4$.
Figure \ref{fig:qg_frac} shows the fractional contributions of the $gb\to \gamma b$ subprocesses
to the total cross section of \gb production
with photons in the central and forward photon rapidity regions as a function of $\Ptg$.
The curves are obtained using signal processes $gb\to \gamma b$ 
and $q\bar{q}\to \gamma b\bar{b}$ simulated
with the {\sc pythia} event generator~\cite{PYT}.
It can be seen that the Compton-like 
contribution is large at small $\ptg$ and decreases with
growing $\ptg$, with the annihilation process contribution having the opposite behavior.

The D0 detector is a general purpose detector 
discussed in detail elsewhere~\cite{d0det}.
The subdetectors most relevant to this analysis are the central tracking
system, composed of a silicon microstrip tracker (SMT) and a central fiber
tracker (CFT) embedded in a 1.9~T solenoidal magnetic field, the central
preshower detector (CPS), and the calorimeter.
The CPS is located immediately before the inner layer of the central calorimeter
and is formed of approximately one radiation length of lead absorber followed by three
layers of scintillating strips. The calorimeter consists of a central section with
coverage in pseudorapidity of $|\eta_{\rm det}|<1.1$~\cite{d0_coordinate}, 
and two end calorimeters covering up to $|\eta_{\rm det}| \approx 4.2$.
The electromagnetic (EM) section of the
calorimeter is segmented longitudinally into four layers (EM$i$, $i=1-4$), 
with transverse segmentation into cells of size 
$\Delta\eta_{\rm det}\times\Delta\phi_{\rm det} = 0.1\times 0.1$~\cite{d0_coordinate}, 
except EM3 (near the EM shower maximum), where it is $0.05\times 0.05$.
The calorimeter allows for a precise measurement of the energy
and direction of electrons and photons,
providing an energy resolution of approximately $4\%$~($3\%$) at an energy of $30 ~(100)$~GeV,
and an angular resolution of  $0.01$ radians.
The energy response of the calorimeter to photons is calibrated using
electrons from $Z$ boson decays. Since electrons and photons shower
differently in matter, additional energy corrections as a function of $y^\gamma$ are
derived using a detailed {\sc geant}-based \cite{Geant} simulation 
of the D0 detector response. These corrections are largest, $\approx 2$\%,
at photon energies of about 30 GeV.
The data used in this analysis 
satisfy D0 data quality requirements and
are collected using a combination of triggers 
requiring a cluster of energy in the electromagnetic (EM) calorimeter with
loose shower shape requirements,
and correspond to an integrated luminosity of $8.7 \pm 0.5$~fb$^{-1}$~\cite{d0lumi}.
The trigger efficiency is $\approx\!96\%$ for photon candidates with 
$p_T^\gamma \sim\!\!30$~GeV and $\approx\!100\%$ for $p_T^\gamma>40$~GeV.

Offline event selection requires a reconstructed $p\bar{p}$ interaction vertex~\cite{pv} 
within 60~cm of the center of the detector along the beam axis. 
The efficiency of the vertex requirement
is $\approx\!(96-98)\%$, depending on $\Ptg$. The missing transverse momentum in the event is required to be less than $0.7
p_{T}^{\gamma}$ to suppress background from 
$W\to e\nu$ decays.  Such a requirement is highly efficient for
signal events, with an efficiency $\geq 98\%$ even for events with
semi-leptonic heavy-flavor quark decays.

To reconstruct photon candidates, projective towers of calorimeter cells with large
deposits of energy are used as seeds to create clusters of energy
in the EM calorimeter in a cone of radius ${\cal R}=0.4$, where 
${\cal R}\equiv\sqrt{(\Delta\eta)^2+(\Delta\phi)^2}$. Once an
EM energy cluster is formed, the final energy ($E_\text{EM}$) is 
obtained summing the energies of all the calorimeter cells in a smaller cone
of ${\cal R}=0.2$.  Photon candidates are required to have: 
(i) $>97$\% of their energy in the EM section; (ii) calorimeter isolation 
$\mathcal{I}=[E_{\text{tot}}(0.4)-E_{\text{EM}}(0.2)]/E_{\text{EM}}(0.2)<0.07$,
where $E_{\text{tot}}({\cal R})$ [$E_{\text{EM}}({\cal R})$] is the total [EM only]
energy in a cone of radius ${\cal R}$; (iii) scalar sum of $p_T$ less than $1.5$~GeV,
calculated from all tracks with $p_T>0.5$ GeV originating from the $p\bar{p}$ primary interaction point 
in an annulus of $0.05<{\cal R}<0.4$ around the EM cluster; and (iv) energy-weighted EM shower width
consistent with that expected for an electromagnetic shower.
To suppress electrons misidentified as photons,
the EM clusters are required to be not spatially matched to significant tracker activity,
either a reconstructed track or, in the central rapidity region,
a density of hits in the SMT and CFT consistent
with that of an electron~\cite{HOR}.
In the following, this requirement is referred to  as the ``track-match veto.''

To further suppress jets misidentified as photons, an artificial neural network ($\gamma$-NN) 
discriminant is defined~\cite{hgg_prl}. It relies on differences between photons and jets
in tracker activity, energy deposits in the calorimeter, and in the CPS for the central photons/jets. 
This $\gamma$-NN is trained using {\sc pythia}~\cite{PYT} Monte Carlo (MC) samples of photon and jet production,
which are processed through a {\sc geant}-based~\cite{Geant} simulation of the
detector geometry and response.
In order to accurately model the effects of multiple $p\bar{p}$ interactions
and detector noise, events from random $p\bar{p}$ 
crossings with a similar instantaneous luminosity spectrum as
in data are overlaid on the MC
events. These MC events are then processed using the
same reconstruction code as for the data.
The $\gamma$-NN performance is verified using a data sample consisting of 
photons radiated from leptons in $Z$ boson decays 
($Z\to\ell^+\ell^-\gamma$, $\ell=e, \mu$)~\cite{Zg}. 
The $\gamma$-NN output $O_{\rm NN}$ distributions for photons in data and MC are in good agreement
and exhibit a significant separation from the distribution
for misidentified jets~\cite{diphoton,hgg_prl}.
Photon candidates are required to have $O_{\rm NN}>0.3$, which is $\approx\!98\%$ efficient for 
photons.

We calculate corrections to the observed number of candidate events
to account for the photon detection efficiency and for the acceptance of the selection using simulated samples
of $\gamma+b$-jet events.
In these samples, the photon is required to be isolated at particle level
by $E_T^{\rm iso} = E_T^{\rm tot}(0.4) - E_T^\gamma < 2.5$ GeV,
 where $E_T^{\rm tot}$ is the total transverse energy of particles within a cone of radius ${\cal R} = 0.4$
centered on the photon, and $E_T^\gamma$ is the photon transverse energy.
Here, the particle level includes all stable particles as defined in Ref.~\cite{particle}.
Signal events are generated using the {\sc sherpa}~\cite{Sherpa} and {\sc pythia} event generators,
processed through a {\sc geant}-based~\cite{Geant} simulation and events reconstruction as described above.
The acceptance 
is driven by the selection requirements in $\eta_{\rm det}$ 
(applied to avoid edge effects in the calorimeter regions used for the measurement) 
and $\phi_{\rm det}$ in the central rapidity region
(to avoid periodic calorimeter module boundaries \cite{d0det} 
that bias the EM cluster energy and position measurements), 
photon rapidity $y^\gamma$ and energy, and bin-to-bin migration effects due to the finite energy and angular resolution of the EM calorimeter.
The acceptance varies within ($82-90$)\% with a relative systematic uncertainty of ($2 - 5$)\%.
The EM clusters reconstructed in the acceptance region are
required to pass the photon identification criteria listed in the previous paragraph.
Average correction factors to account
for differences between data and simulations are obtained with
the  {\sc sherpa} events, while the difference from the corrections 
obtained with {\sc pythia} is used as systematic uncertainty.
Small differences between data and MC in the photon selection efficiencies are corrected for with suitable scale factors
derived using control samples of electrons from $Z$ boson decays, as well as photons from
the radiative $Z$ boson decays~\cite{Zg}.
The total efficiency of the above photon selection criteria
is ($68-85$)\%, depending on the \ptg and rapidity region.  
The systematic uncertainties on these values are 3\% for $|y^\gamma|<1.0$ 
and $7.3\%$ for $1.5<|y^\gamma|<2.5$ and are mainly due to uncertainties 
caused by the track-match veto, isolation, and the $\gamma$-NN requirements.
The contamination from $Z(\to\!e^+e^-)$+jet and $W(\to\!e\nu)$+jet events
is estimated from the simulation and is found to be negligible ($\lesssim1\%$) 
for both photon rapidity regions.

At least one jet with $p_{T}^{\rm jet}>15$~GeV and $| y^\text{jet}|\lt 1.5$ 
must be present in each selected event. Jets are reconstructed
using the D0 Run~II algorithm~\cite{Run2Cone} with a cone radius of $\mathcal{R}=0.5$.  
The jet acceptance with respect to the $p_{T}^{\rm jet}$ and $|y^\text{jet}|$ kinematic cuts
varies between $88\%$ and $100\%$ in different photon $p_T$ bins.
The uncertainties on the acceptance due to the jet energy scale, jet energy
resolution, and difference in energy scale correction between light flavor and $b$-jets 
vary between $1$\% and $7$\%, increasing for smaller $\Ptg$.
The jet is required to have at least two associated tracks with $p_T>0.5$~\GeV
with at least one hit in the SMT.
The track with the highest $p_T$ must have $p_T>1.0$~\GeV.
These criteria ensure that there is  sufficient information to 
classify the jet as a heavy-flavor candidate and have a typical efficiency 
of about 90\%.  Light jets (caused by light quarks or gluons) are suppressed
using a dedicated artificial neural network ($b$-NN)~\cite{b-NIM}
that employs the longer lifetimes of heavy-flavor hadrons relative to
their lighter counterparts. 
The inputs to the $b$-NN combine several characteristic quantities of the
jet and associated tracks to provide a continuous output
value that tends towards one for $b$-jets and zero for the light
jets. The $b$-NN input variables providing most of the discrimination are
the number of 
reconstructed secondary vertices (SV) in the jet, the invariant mass of charged
particle tracks associated with the SV ($M_{\rm SV}$), 
the number of tracks used to reconstruct the SV, the two-dimensional
decay length significance of the SV in the plane transverse
to the beam, a weighted combination of the tracks' transverse impact parameter significances, 
and the probability that the tracks associated with the jet originate from the $p\bar{p}$ interaction vertex.
The jet is required to have a 
$b$-NN output $>0.3$. 
Depending on $\ptg$, this selection is ($40-52$)\% efficient for $b$-jets
with systematic uncertainties of
($6-23$)\% for the \gb events with $|y^\gamma|<1.0$
and of ($7-11$)\% for those with $1.5<|y^\gamma|<2.5$, both increasing as a function of $\Ptg$.
Only $0.2-0.4$\% of light jets are misidentified as heavy-flavor jets, comprising 
$7$\% to $10$\% of the final sample.

\begin{figure*}
\includegraphics[width=0.45\linewidth]{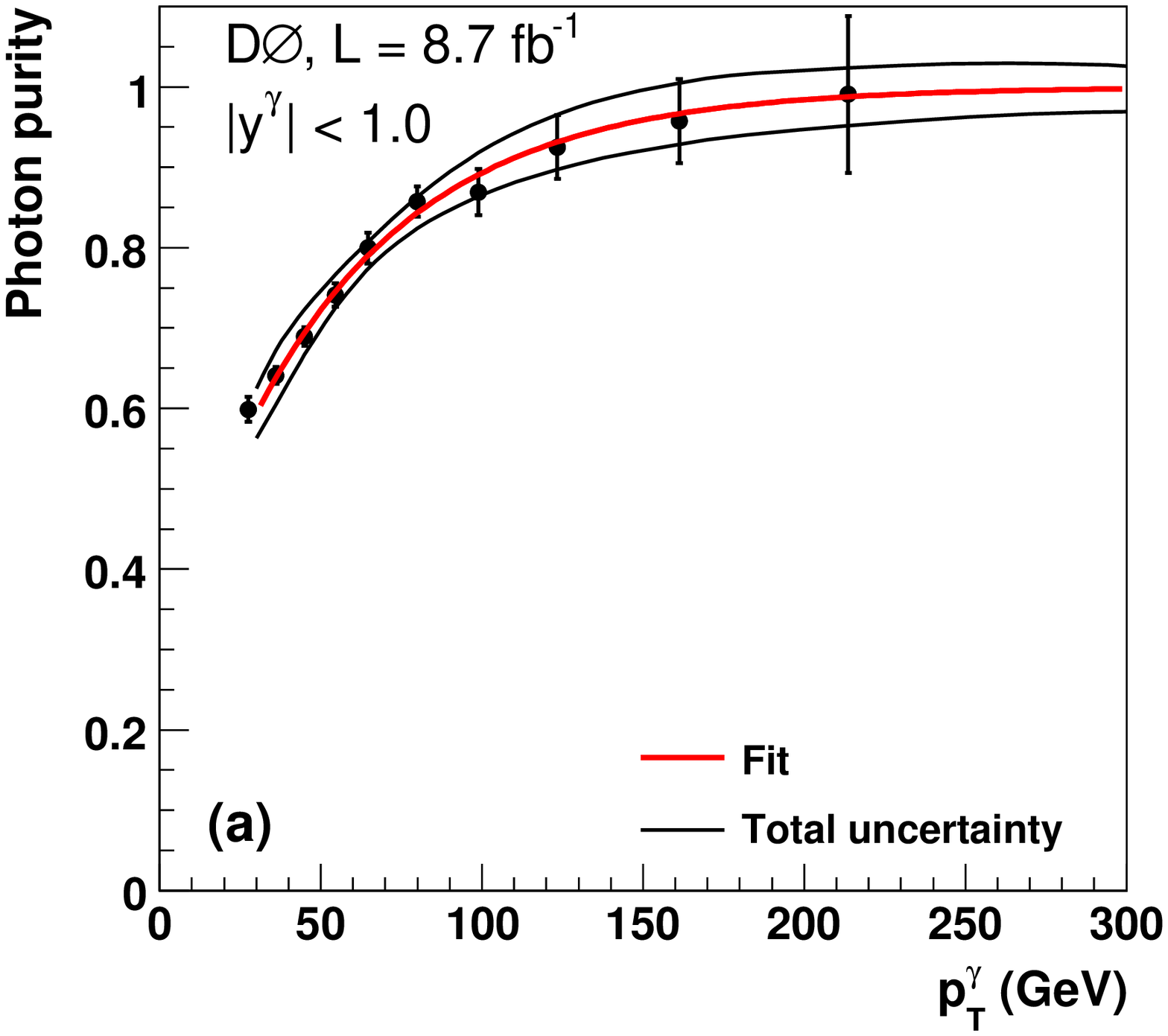} 
\includegraphics[width=0.45\linewidth]{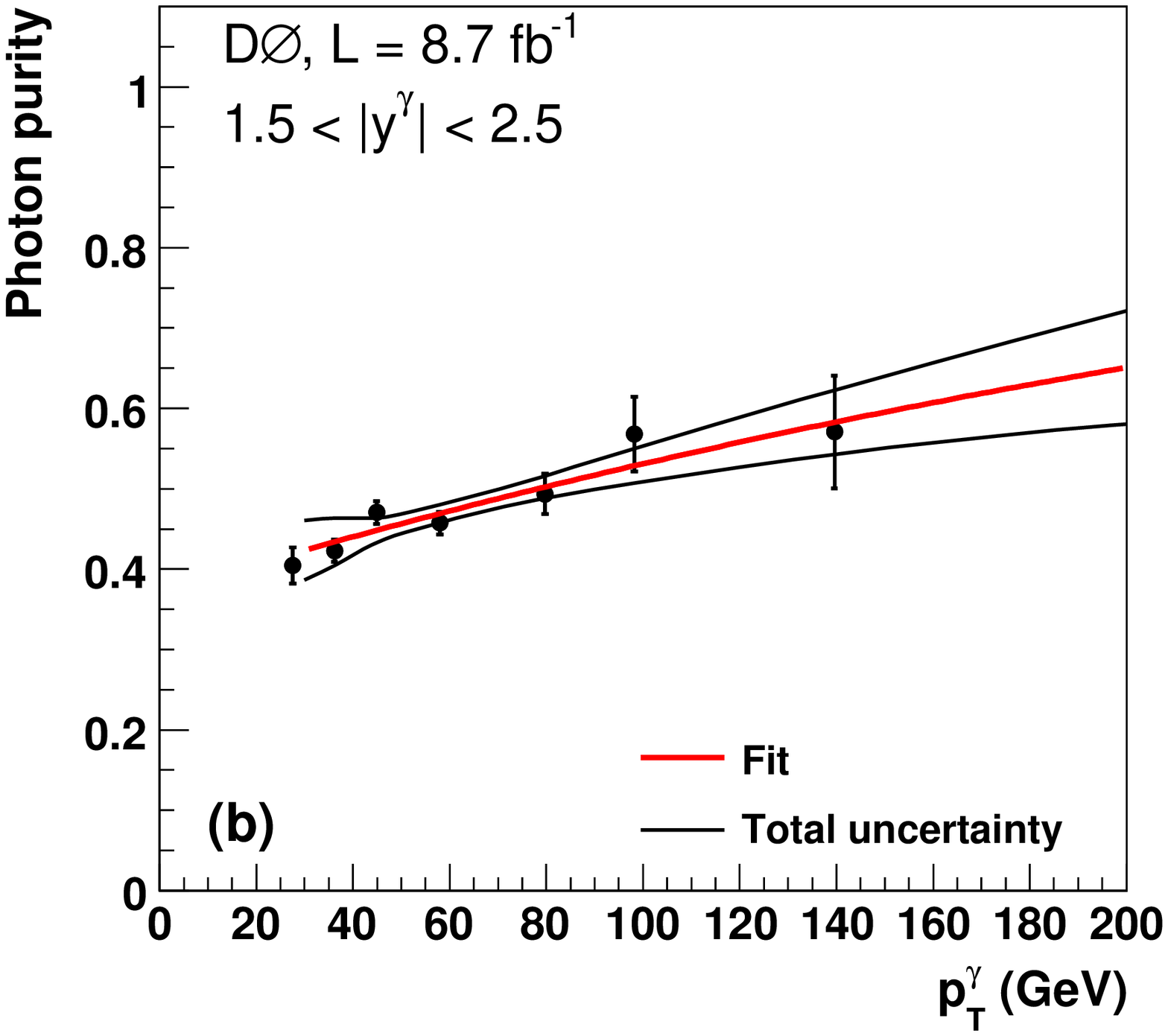}
~\\[-5mm]
\caption{Photon purity as a function of $\Ptg$ in the selected data sample 
in the central rapidity region $|y^\gamma|<1.0$ (a) and the forward rapidity region $1.5<|y^\gamma|<2.5$ (b).}
\label{fig:gam_pur_cc}
\end{figure*}

After all selection requirements, 
199,515 (139,710) events remain in the data samples
with the central (forward) photons.  
In addition to events with light-flavor jets a second source of background
is represented by multi-jet events in which one jet is  misidentified as a photon.
To estimate the
photon purity, the $\gamma$-NN distribution in data is fitted to a
linear combination of templates for photons and jets obtained from
simulated $\gamma~+$ jet and dijet samples, respectively.  An
independent fit is performed in each $\Ptg$ bin, yielding photon
purities between 62\% and 99\% for the events
with the central photons and between 40\% and 55\% for the events
with the forward photons. The obtained photon fractions are 
shown in Fig.~\ref{fig:gam_pur_cc}. 
The  $\Ptg$ dependence of the purity is fitted in each region using 
a two-parameter function ${\cal P}=1-\exp(a+b\Ptg)$.
The systematic uncertainties on the fit are estimated using two alternative
fitting functions. 
These photon purities differ at most by $7$\% when compared with those obtained for inclusive
events, i.e. without the requirement of a heavy-flavor jet.
An additional systematic uncertainty in the photon fractions 
due to the fragmentation model implemented in {\sc pythia} is also taken into account~\cite{PhotonInc}.
This uncertainty is estimated by varying the
production rate of $\pi^0$ and $\eta$ mesons by $\pm$50\% with respect to their
central values~\cite{Frag}.
It is found to be about $6\%$ at $\Ptg\simeq 30$ GeV, $2\%$ at $\Ptg\simeq 50$ GeV, 
and $\leq1\%$ at $\Ptg\gtrsim 70$ GeV.

The fractions of $b$-jets are determined by fitting $M_{\rm SV}$
templates for $b,c$ and light jets to the data. 
Jets from $b$-quarks tend to have larger values of $M_{\rm SV}$,
in contrast to light jets.
For $b$- and $c$-jets, the templates are
obtained from simulation, while the light jet template is derived from a data sample, 
enriched in light jets, referred to as negatively tagged data (NT data). 
The NT data comprises the jets that have negative 
values for some of the inputs to the $b$-NN  algorithm 
(such as negative decay length and negative impact parameter significance) which are caused
by detector resolution effects~\cite{b-NIM}.
After correcting the NT data for the small contamination from heavy-flavor jets,
we have verified that the $M_{\rm SV}$ template shapes in NT data and light jets in the MC simulation
agree well. 

\begin{figure}
\includegraphics[width=0.85\linewidth]{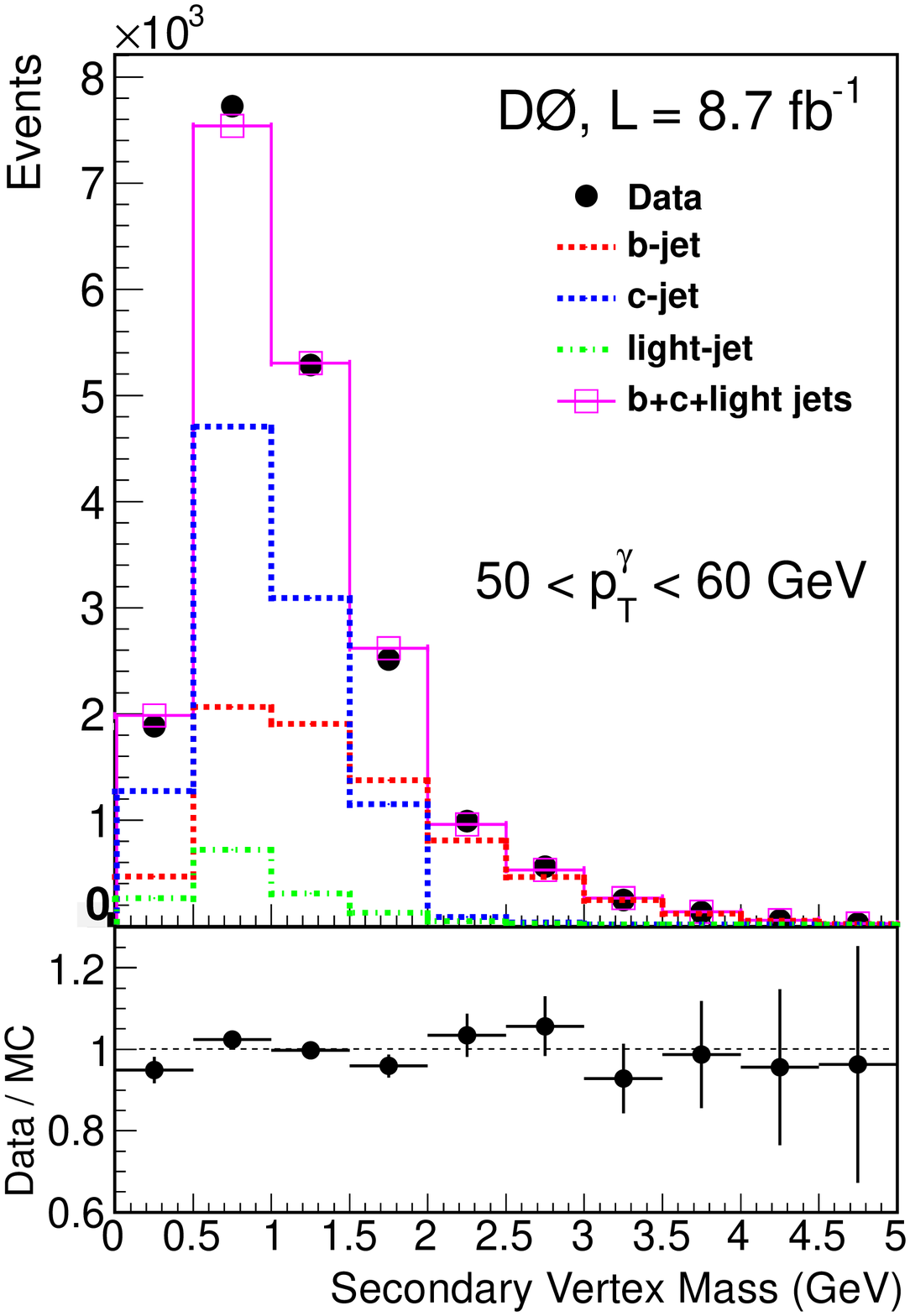} 
\caption{Distribution of observed events for secondary vertex mass
  after all selection criteria for the representative bin $50<\Ptg< 60$~GeV ($|y^\gamma|<1.0$).  
  The distributions for the $b$-, $c$-, and light jet templates are shown
  normalized to their respective fitted fractions.
 Also included at the bottom is the ratio of data to the result of the fit.  
   Fits in the other \ptg bins are of similar quality.
  }
\label{fig:cbjet_test}
\end{figure}

\begin{figure*}
\includegraphics[width=0.45\linewidth]{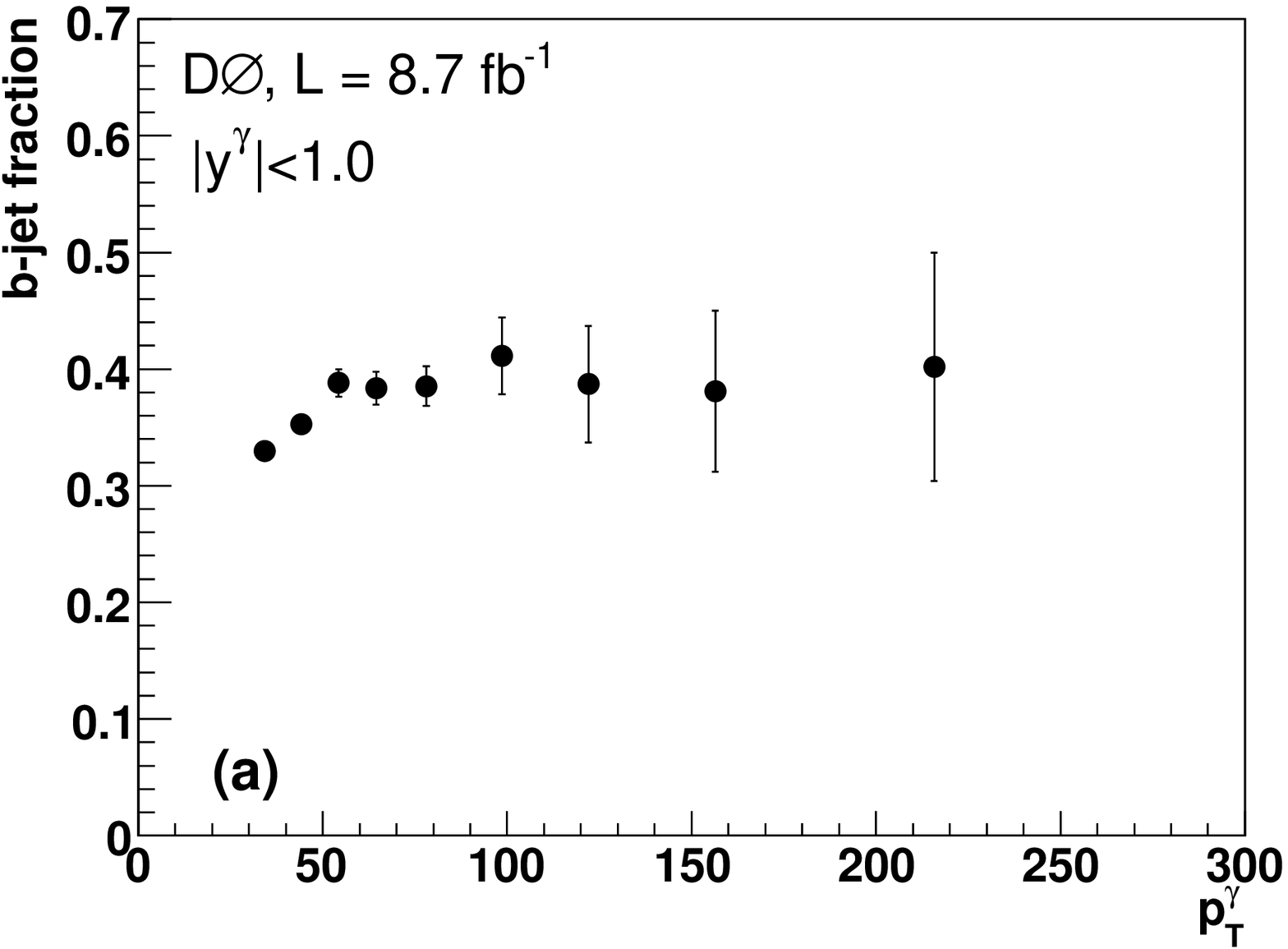} 
\includegraphics[width=0.45\linewidth]{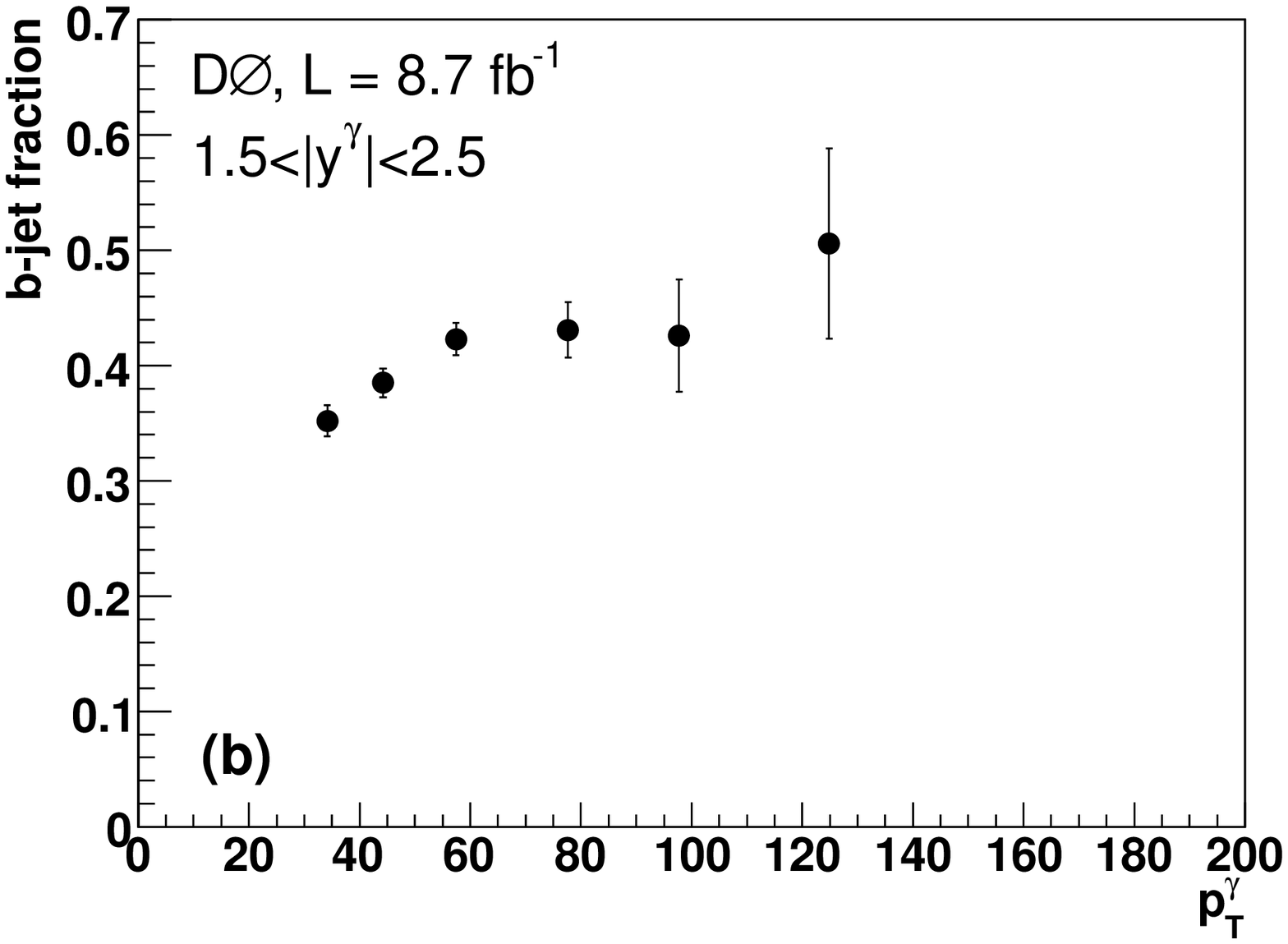}
\caption{The $b$-jet fraction (with total uncertainties) as a function of $\Ptg$ in the data sample after applying all selections
in the central rapidity region $|y^\gamma|<1.0$ (a)
and the forward rapidity region $1.5<|y^\gamma|<2.5$ (b).}
\label{fig:b_pur_cc}
\end{figure*}

The result of a maximum likelihood fit to $M_{\rm SV}$ templates, normalized to the number of
events in data, is shown in Fig.~\ref{fig:cbjet_test} for central photons
with $50<\Ptg<60~\GeV$, as an example.  As  shown in Fig.~\ref{fig:b_pur_cc},
the estimated fraction of $b$-jets grows
with $\Ptg$ from about 35\% to about 42\%. 
The corresponding relative uncertainties range between
(4--24)\%, increasing at higher $\Ptg$ and are dominated by the limited data statistics.
The data corrected for the photon and jet acceptance, for reconstruction efficiencies,
for the contribution of background events, and for bin-migration effects, are presented at the particle level,
as defined in Ref.~\cite{particle}.

\begin{figure}
\includegraphics[width=0.99\linewidth]{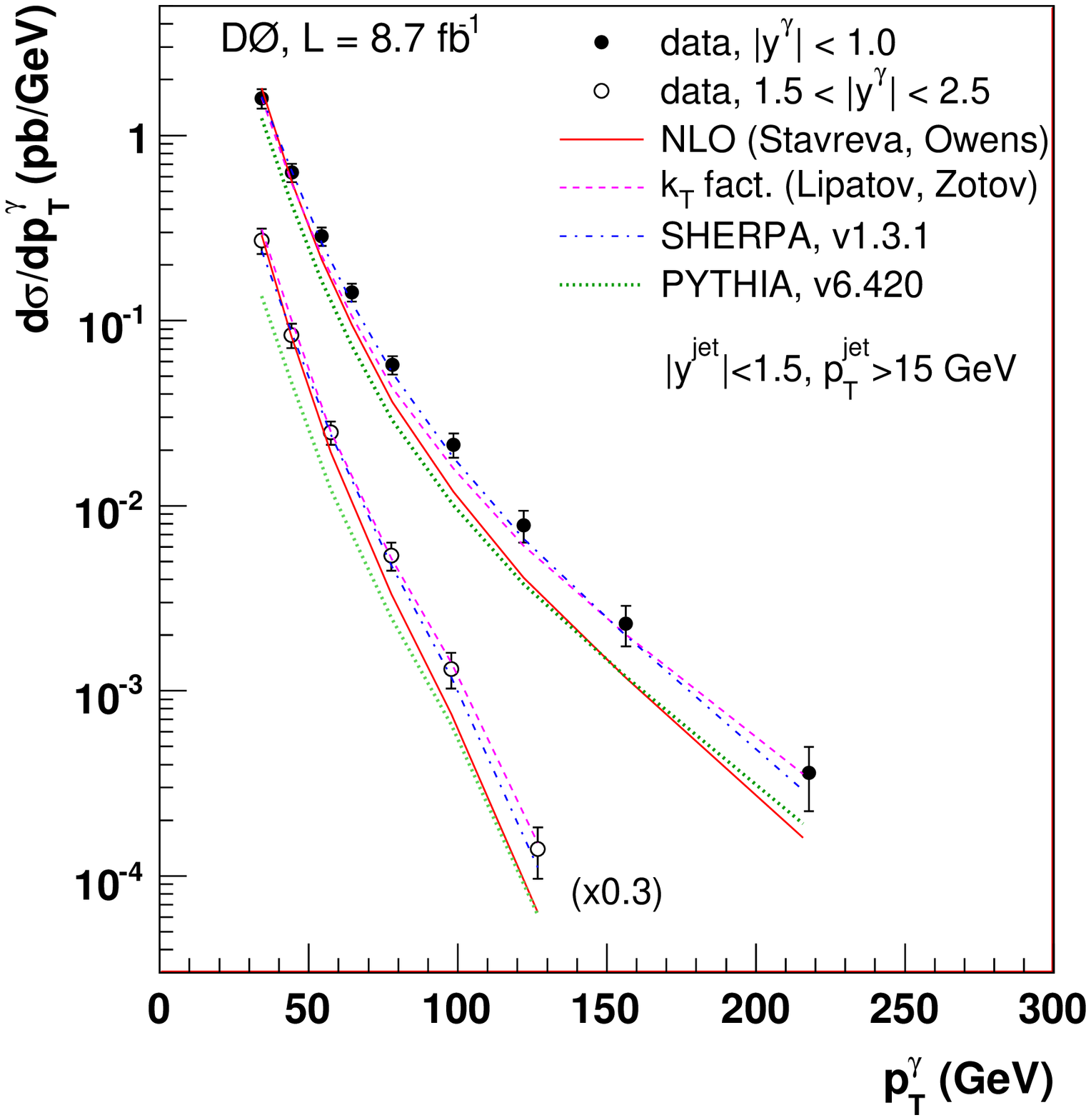}
~\\[-7mm]
\caption{ The \gb production differential cross sections
  as a function of $\Ptg$ in the two photon rapidity regions, $|y^{\gamma}|<1.0$
  and $1.5<|y^{\gamma}|<2.5$ (the latter results are multiplied by 0.3 for presentation).
  The uncertainties on the data points include statistical and systematic contributions
  added in quadrature. The measurements are compared to the NLO QCD calculations using {\sc cteq}6.6M
  PDFs~\cite{CTEQ} (solid line). The predictions from {\sc sherpa}, {\sc pythia} and ``$k_{\rm T}$ factorization'' approach~\cite{Zotov,Zotov2} 
  are shown by the dash-dotted, dotted and dashed lines, respectively.}
\label{fig:xsectLOGplot}
\end{figure}

The differential cross sections of \gb production are extracted in nine (six) bins of \ptg{}
for central (forward) photons, and are listed
in Tables~\ref{tab:xsect_cc} and \ref{tab:xsect_ec}.  
The results are also shown in Fig.~\ref{fig:xsectLOGplot}
as a function of $\Ptg$ for the two photon rapidity intervals. 
The data points are plotted at the value of $\Ptg$ for which
the value of the smooth function describing the cross section 
equals the averaged cross section in the bin~\cite{TW}.
The $\gamma~+b$-jet simulated sample with Sherpa has been used to determine mean $\Ptg$ values.

The cross sections with the central (forward) photons fall by
about four (three) orders of magnitude in the range $30<\Ptg< 300~(200)~\GeV$.
The statistical uncertainty on the results ranges from 2\% in the
first $\Ptg$ bin to $\approx 11\%$ in the last $\Ptg$ bins, while the total
systematic uncertainty varies between 12\% and 36\%.  The main source of the
uncertainty at low \ptg{} is due to the photon purity (up to $8\%$), the
$b$-jet fraction fit ($6-7\%$), and the luminosity (6.1\%)~\cite{d0lumi}.  
At higher \ptg{}, the uncertainty is
dominated by the fractions of $b$-jets and their selection efficiencies.  
Systematic uncertainties are highly bin-to-bin correlated for the first three $\Ptg$ bins,
while the total systematic uncertainty is nearly uncorrelated
across the bins at $\Ptg>70$~\GeV.

Next-to-leading order (NLO) perturbative QCD 
 predictions, with
the renormalization scale $\mu_{R}$, factorization scale $\mu_{F}$,
and fragmentation scale $\mu_f$ all set to $\Ptg$, are also given in
Tables \ref{tab:xsect_cc} and \ref{tab:xsect_ec}.
These predictions~\cite{Tzvet} are
based on a phase space slicing method used to calculate the cross section analytically~\cite{Harris}.  
The uncertainty from the choice of scale is estimated 
through a simultaneous variation of all three scales by a factor of two, i.e.
for $\mu_{R,F,f}=0.5 p_T^\gamma$ and $2 p_T^\gamma$. The predictions
utilize {\sc cteq}6.6M PDFs~\cite{CTEQ} and are corrected for non-perturbative effects
of parton-to-hadron fragmentation and multiple parton interactions. 
The latter are evaluated using {\sc sherpa} and {\sc pythia} MC samples using their default
settings~\cite {Sherpa,PYT}.
The overall correction varies from about $0.90$ at $30<\Ptg<40~\GeV$ to about $0.95$ at high $\Ptg$,
and an uncertainty of $\lesssim 2\%$ is assigned to account for the
difference between the two MC generators.

\begin{figure*}
\includegraphics[width=0.49\linewidth]{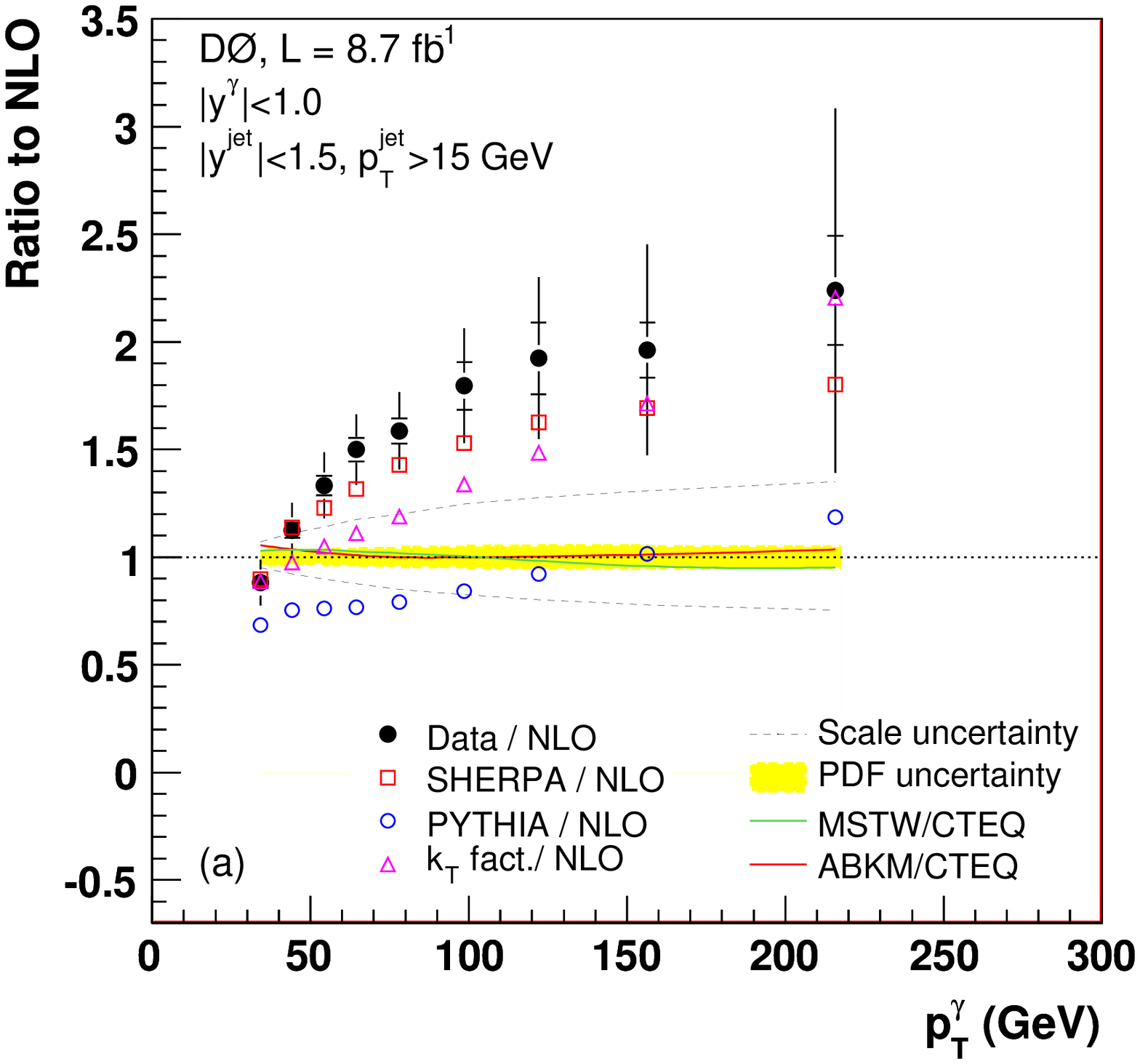}
\includegraphics[width=0.49\linewidth]{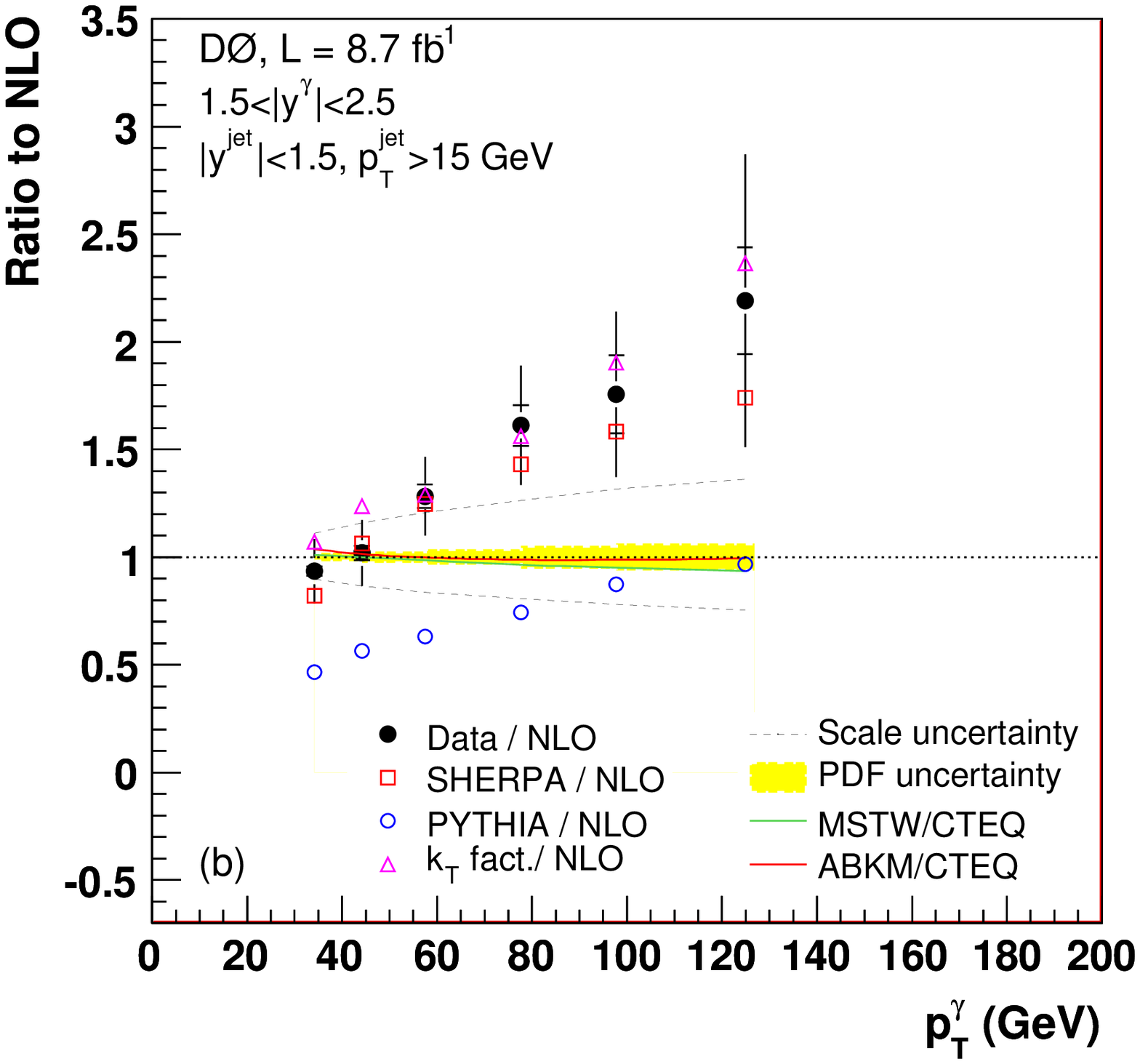}
\caption{
The ratio of \gb production differential cross sections between data and NLO QCD predictions with uncertainties
for the rapidity regions $|y^{\gamma}|<1.0$ (a) and $1.5<|y^{\gamma}|<2.5$ (b).
The uncertainties on the data include both statistical (inner error bar) and full uncertainties (entire error bar).  
Also shown are the uncertainties on the theoretical QCD scales and the {\sc cteq}6.6M PDFs. The ratio of NLO predictions with {\sc cteq}6.6M to those with 
{\sc mstw2008}~\cite{mstw} and {\sc abkm09nlo}~\cite{abkm} are also shown.
}
\label{fig:xsectratio}
\end{figure*}

The prediction based on the $k_T$-factorization approach~\cite{Zotov,Zotov2} 
and unintegrated parton distributions~\cite{UPD}
are also given in Tables \ref{tab:xsect_cc} and \ref{tab:xsect_ec}.
The $k_T$-factorization formalism contains additional contributions 
to the cross sections due to resummation of
gluon radiation diagrams with $k_T^2$ above a scale $\mu^2$, $O(1~ \rm{GeV})$, 
where  $k_T$ denotes the transverse momentum of the radiated gluon. 
Apart from this resummation, the non-vanishing transverse momentum 
distribution of the colliding partons are taken into account. 
These effects lead to a broadening of the photon transverse momentum distribution 
in this approach~\cite{Zotov}.
The scale uncertainties on these predictions 
vary from about $31\%$ ($-28\%$) at $30<\Ptg<40~\GeV$ to about 
5\% (14\%) for the central photons and $26\%$ ($-13\%$) for the forward photons
in the last $\Ptg$ bin.

Tables \ref{tab:xsect_cc} and \ref{tab:xsect_ec} also contain predictions 
of the {\sc pythia} \cite{PYT} MC event generator with the {\sc cteq}6.1L PDF set.
It includes only $2\to 2$ matrix elements (ME) with $gb\to \gamma b$ and $q\bar{q}\to \gamma g$ scatterings 
(defined at LO) and, with $g\to b\bar{b}$ splitting in the parton shower (PS).
We also provide predictions by the
{\sc sherpa} MC event generator \cite{Sherpa} with the {\sc cteq}6.6M  PDF set~\cite{CTEQ}.
For \gb production, {\sc sherpa}
includes all the MEs with one photon and up to three
jets, with at least one $b$-jet in our kinematic region. 
In particular, 
it accounts for an additional hard jet that accompanies the photon 
associated with a $b\bar{b}$ pair. Compared to an  NLO calculation,
there is an additional benefit of imposing resummation (further emissions)
through the consistent combination with the PS.
Matching between the ME partons and the PS jets follows the
prescription given in Ref.~\cite{SHERPA_gam}.
Systematic uncertainties are estimated by varying the ME-PS matching scale by $\pm 5$ GeV
around the chosen central value \cite{Sherpa_scale}.
As a result, the {\sc sherpa} cross sections vary up to $\pm7\%$ for the central photons
and up to $-25\%$/$+17\%$ for the forward photons, where the uncertainty is largest in the first $\Ptg$ bin.

 All the theoretical predictions are obtained including the isolation requirement 
on the photons $E_T^{\rm iso}<2.5$ GeV at the particle level.
The predictions are compared to data in Fig.~\ref{fig:xsectLOGplot} as a function of $\Ptg$.
The ratios of data over the NLO QCD calculations and of different QCD predictions or MC simulation to
the same NLO QCD calculations are shown in Fig.~\ref{fig:xsectratio} as a function of $\Ptg$ for both central and forward photons.
The ratio of NLO predictions with {\sc cteq}6.6M to those with
{\sc mstw2008}~\cite{mstw} and {\sc abkm09nlo}~\cite{abkm} are also shown on the plots.

\begin{table*}
\centering
\caption{The \gb production cross section ${\rm d}\sigma/{\rm d}\Ptg$ in bins of $\Ptg$ for $|y^\gamma|<1.0$
together with statistical, $\delta\sigma_{\text{stat}}$,  and systematic, $\delta\sigma_{\text{syst}}$, uncertainties.
The last four columns show theoretical predictions obtained within NLO QCD, k$_{\rm T}$ factorization, 
{\sc pythia} and {\sc sherpa} event generators.}
\label{tab:xsect_cc}
\begin{tabular}{cccccccccc} \hline 
 ~$\Ptg$ bin~ & ~$\la\Ptg\ra$~ & \multicolumn{8}{c}{${\rm d}\sigma/{\rm d}\Ptg$ (pb/GeV) } \\\cline{3-10}
  (GeV) & (GeV) & Data & $\delta_{\rm stat}$($\%$) & $\delta_{\rm syst}$($\%$) & $\delta_{\rm tot}$($\%$) & ~~~NLO~~~ & ~~~k$_{\rm T}$ fact.~~~ & ~~~{\sc pythia}~~~ & ~~~{\sc sherpa}~~~\\\hline  
   30 --  40 &  34.2 &  1.59$\times 10^{0}$~~ &    2 &   12 &   12 &   1.80$\times 10^{0}$~~ &  1.60$\times 10^{0}$~~ &  1.24$\times 10^{0}$~~ &  1.62$\times 10^{0}$~~ \\\hline
   40 --  50 &  44.3 &  6.30$\times 10^{-1}$ &    3 &   11 &   11 &   5.60$\times 10^{-1}$ &  5.47$\times 10^{-1}$ &  4.23$\times 10^{-1}$ &  6.38$\times 10^{-1}$ \\\hline
   50 --  60 &  54.3 &  2.85$\times 10^{-1}$ &    3 &   11 &   11 &   2.14$\times 10^{-1}$ &  2.25$\times 10^{-1}$ &  1.63$\times 10^{-1}$ &  2.63$\times 10^{-1}$ \\\hline
   60 --  70 &  64.5 &  1.42$\times 10^{-1}$ &    4 &   10 &   11 &   9.49$\times 10^{-2}$ &  1.05$\times 10^{-1}$ &  7.27$\times 10^{-2}$ &  1.25$\times 10^{-1}$ \\\hline
   70 --  90 &  78.1 &  5.77$\times 10^{-2}$ &    4 &   11 &   11 &   3.64$\times 10^{-2}$ &  4.32$\times 10^{-2}$ &  2.88$\times 10^{-2}$ &  5.20$\times 10^{-2}$ \\\hline
   90 -- 110 &  98.6 &  2.14$\times 10^{-2}$ &    6 &   14 &   15 &   1.19$\times 10^{-2}$ &  1.59$\times 10^{-2}$ &  1.00$\times 10^{-2}$ &  1.82$\times 10^{-2}$ \\\hline
  110 -- 140 & 122.0 &  7.85$\times 10^{-3}$ &    9 &   18 &   20 &   4.08$\times 10^{-3}$ &  6.06$\times 10^{-3}$ &  3.76$\times 10^{-3}$ &  6.63$\times 10^{-3}$ \\\hline
  140 -- 180 & 156.4 &  2.31$\times 10^{-3}$ &    7 &   24 &   25 &   1.18$\times 10^{-3}$ &  2.02$\times 10^{-3}$ &  1.19$\times 10^{-3}$ &  1.99$\times 10^{-3}$ \\\hline
  180 -- 300 & 215.8 &  3.60$\times 10^{-4}$ &   11 &   36 &   38 &   1.61$\times 10^{-4}$ &  3.55$\times 10^{-4}$ &  1.91$\times 10^{-4}$ &  2.90$\times 10^{-4}$ \\\hline
\end{tabular}
\end{table*}
\vspace*{6ex}
\begin{table*}
\centering
\caption{The \gb production cross section ${\rm d}\sigma/{\rm d}\Ptg$ in bins of $\Ptg$ for $1.5<|y^\gamma|<2.5$
together with statistical, $\delta\sigma_{\text{stat}}$,  and systematic, $\delta\sigma_{\text{syst}}$, uncertainties.
The last four columns show theoretical predictions obtained within NLO QCD, k$_{\rm T}$ factorization, {\sc pythia} and {\sc sherpa} event generators.}
\label{tab:xsect_ec}
\begin{tabular}{cccccccccc} \hline 
 ~$\Ptg$ bin~ & ~$\la\Ptg\ra$~ & \multicolumn{8}{c}{${\rm d}\sigma/{\rm d}\Ptg$ (pb/GeV) } \\\cline{3-10}
  (GeV) & (GeV) & Data & $\delta_{\rm stat}$($\%$) & $\delta_{\rm syst}$($\%$) & $\delta_{\rm tot}$($\%$) & ~~~NLO~~~ & ~~~k$_{\rm T}$ fact.~~~ & ~~~{\sc pythia}~~~ & ~~~{\sc sherpa}~~~\\\hline  
   30 --  40 &  34.2 &  9.05$\times 10^{-1}$ &    2 &   16 &   16 &   9.68$\times 10^{-1}$ &  1.04$\times 10^{0}$~~ &  4.51$\times 10^{-1}$ &  7.95$\times 10^{-1}$ \\\hline
   40 --  50 &  44.2 &  2.79$\times 10^{-1}$ &    3 &   15 &   15 &   2.74$\times 10^{-1}$ &  3.38$\times 10^{-1}$ &  1.55$\times 10^{-1}$ &  2.91$\times 10^{-1}$ \\\hline
   50 --  70 &  57.4 &  8.30$\times 10^{-2}$ &    4 &   14 &   14 &   6.47$\times 10^{-2}$ &  8.34$\times 10^{-2}$ &  4.08$\times 10^{-2}$ &  8.06$\times 10^{-2}$ \\\hline
   70 --  90 &  77.7 &  1.79$\times 10^{-2}$ &    6 &   16 &   17 &   1.11$\times 10^{-2}$ &  1.74$\times 10^{-2}$ &  8.26$\times 10^{-3}$ &  1.59$\times 10^{-2}$ \\\hline
   90 -- 110 &  97.8 &  4.38$\times 10^{-3}$ &   10 &   19 &   22 &   2.50$\times 10^{-3}$ &  4.75$\times 10^{-3}$ &  2.18$\times 10^{-3}$ &  3.95$\times 10^{-3}$ \\\hline
  110 -- 200 & 124.9 &  4.65$\times 10^{-4}$ &   11 &   29 &   31 &   2.12$\times 10^{-4}$ &  5.02$\times 10^{-4}$ &  2.06$\times 10^{-4}$ &  3.70$\times 10^{-4}$ \\\hline
\end{tabular}
\end{table*}

The measured cross sections are in agreement with the NLO QCD predictions 
within theoretical and experimental uncertainties 
in the region up to $\Ptg \lesssim 70$ GeV, but show notable disagreement
for larger $\Ptg$. The cross section slopes in data 
significantly differ from the predicted ones in both photon rapidity regions.
The results indicate a need for higher order perturbative QCD corrections
in the large $\Ptg$ region, that is dominated by the annihilation process $q\bar{q}\to \gamma g$ ($g\to b\bar{b}$),
and resummation of diagrams with additional gluon radiation. 
The QCD predictions from the $k_T$-factorization approach is in better agreement with data.
The best agreement is obtained with the {\sc sherpa} MC that
allows up to two extra hard partons (jets) in ME in addition to the $b$-quark (jet) and includes
the consistent treatment of the possible contributions from the parton shower.
However, correcting the {\sc sherpa} predictions by including
additional higher order contributions, as is done, for instance, for $W+$jets events,
would be desirable~\cite{Blackhat}.

In conclusion, we have performed a measurement of the
differential cross section of inclusive production of photon in
association with $b$-jets at the Tevatron $p\bar{p}$ collider. 
The results cover the kinematic ranges $30<\ptg<300~\GeV$ with 
$|y^\gamma|<1.0$, and $30<\ptg<200~\GeV$ with 
$1.5<|y^\gamma|<2.5$. 
A good description of the data can only be achieved by including higher order corrections
into the NLO QCD predictions, which are currently present as additional real emissions
in the {\sc sherpa} MC generator. 
These results can be used to improve the description of background processes 
in searches for the Higgs boson or for new phenomena beyond the SM 
at the Tevatron and the LHC 
in final states involving the production of vector bosons in association with $b$-jets.
  
We are grateful to the authors of the theoretical calculations,
T.~Stavreva, J.~Owens, N.~Zotov, S.~Schumann and F.~Siegert, 
for providing predictions and for many useful discussions.

%
We thank the staffs at Fermilab and collaborating institutions,
and acknowledge support from the
DOE and NSF (USA);
CEA and CNRS/IN2P3 (France);
MON, Rosatom and RFBR (Russia);
CNPq, FAPERJ, FAPESP and FUNDUNESP (Brazil);
DAE and DST (India);
Colciencias (Colombia);
CONACyT (Mexico);
NRF (Korea);
FOM (The Netherlands);
STFC and the Royal Society (United Kingdom);
MSMT and GACR (Czech Republic);
BMBF and DFG (Germany);
SFI (Ireland);
The Swedish Research Council (Sweden);
and
CAS and CNSF (China).

\bibliography{paper_gb}

\begin{thebibliography}{99}
\bibitem{d0_coordinate}
The polar angle $\theta$ and the azimuthal angle $\phi$ are defined
with respect to the positive $z$ axis, which is along the proton beam direction.
Pseudorapidity is defined as $\eta=-\ln[\tan(\theta/2)]$.
Also, $\eta_{\text {det}}$ and $\phi_{\text {det}}$ are the pseudorapidity and the azimuthal angle
measured with respect to the center of the detector.
\bibitem{pv}
The primary $p\bar{p}$ interaction vertex is that found to be the most likely collision point,
among possibly several collisions within a specific beam crossing, from which our selected objects
emanate. The algorithm for defining it can be found in [9].
\bibitem{Sherpa_scale}
We  choose the following ME-PS matching parameters: the energy scale 
$Q_0 = 15$ GeV and the spatial scale $D = 0.4$, where $D$ is taken to
be of the size of the photon isolation cone.
\end{thebibliography}
\bibliographystyle{apsrev}

\end{document}